\documentclass[twocolumn,preprintnumbers,superscriptaddress,nofootinbib,showpacs,showkeys]{revtex4}
\usepackage{amsmath}
\usepackage{amssymb}
\usepackage{graphicx}
\usepackage{hyperref}
\usepackage{xspace}
\usepackage{bm}
\usepackage{slashed}
\usepackage{color}

\usepackage{booktabs}
\AtBeginDocument{
\heavyrulewidth=.08em
\lightrulewidth=.05em
\cmidrulewidth=.03em
\belowrulesep=.65ex
\belowbottomsep=0pt
\aboverulesep=.4ex
\abovetopsep=0pt
\cmidrulesep=\doublerulesep
\cmidrulekern=.5em
\defaultaddspace=.5em
}

\newcommand{\be}{\begin{equation}}
\newcommand{\ee}{\end{equation}}
\newcommand{\ba}{\begin{eqnarray}}
\newcommand{\ea}{\end{eqnarray}}

\newcommand{\helium}{\ensuremath{^3\text{He}}\xspace}
\newcommand{\hypertriton}{\ensuremath{{}_\Lambda^3\text{H}}\xspace}

\begin{document}

\title{The observation of light nuclei at ALICE and the $X(3872)$ conundrum}


\newcommand{\sapienza}{Dipartimento di Fisica and INFN, `Sapienza' Universit\`a di Roma\\
P.le Aldo Moro 5, I-00185 Roma, Italy}
\newcommand{\columbia}{Department of Physics, 538W 120th Street,
Columbia University, New York, NY, 10027, USA}
\newcommand{\pavia}{INFN Pavia, Via A. Bassi 6, I-27100 Pavia, Italy}
\newcommand{\romadue}{Dipartimento di Fisica and INFN, Universit\`a di Roma `Tor Vergata'\\
 Via della Ricerca Scientifica 1, I-00133 Roma, Italy}
\newcommand{\alice}{ALICE\xspace}

\author{A.~Esposito}
\affiliation{\columbia}
\author{A.L.~Guerrieri}
\affiliation{\romadue}
\author{L.~Maiani}
\affiliation{\sapienza}
\author{F.~Piccinini}
\affiliation{\pavia}
\author{A.~Pilloni}
\affiliation{\sapienza}
\author{A.D.~Polosa}
\affiliation{\sapienza}
\author{V.~Riquer}
\affiliation{\sapienza}

\begin{abstract}
The new  data reported by ALICE on the production of light nuclei with $p_\perp\lesssim 10$~GeV in Pb-Pb collisions 
at $\sqrt{s_\text{NN}}=2.76$~TeV are used to compute an order-of-magnitude estimate of the expected production 
cross sections of light nuclei in proton-proton collisions at high transverse momenta.
 We compare the hypertriton, helium-3 and deuteron production cross sections to that of  $X(3872)$,  measured in 
prompt $pp$ collisions by CMS. The results we find suggest a different production mechanism for the $X(3872)$, 
making questionable any loosely bound molecule interpretation. 
 \end{abstract}


\maketitle
As first discussed in~\cite{bigna}, one expects  a suppression of loosely bound hadron molecules in high energy 
$pp(\bar p)$ collisions. Small relative momenta in the center of mass of such molecular hadrons, needed to preserve 
a state with few keVs' binding energies, are in fact hard to be obtained in hadron collisions at high energy and 
$p_\perp$.

Despite this, the $X(3872)$, one of the most studied loosely bound hadron molecule candidates~\cite{moles}, is strongly produced 
at the LHC -- see {\it e.g.}~\cite{cms}. This might simply be the indication that the $X$ hadron molecule 
interpretation is not correct (for the alternative tetraquark model, see~\cite{mainoi,lebed,review}).

Assuming that final state interaction mechanisms are at work -- whose description requires several model-dependent 
assumptions~\cite{fsi,guo} -- it has been proposed  that the relative kinetic energy might be reduced in the 
center of mass of the hadron pair constituting the $X$, in such a way to match a shallow discrete level of some 
inter-hadron potential. A hadron molecule would then be formed, with a precise relation between binding energy and 
strong coupling to its constituent hadrons~\cite{ng}. Since the mass and branching ratios of the $X$ have not been 
measured with the required precision yet,  it is still unclear if this relation is fulfilled. 

Final state interactions should also favor the prompt formation of {\it bona fide} light nuclei in high energy 
hadronic collisions. 
It would therefore be of great interest to measure the $pp$  (anti)deuteron production cross section in the same 
$p_\perp$ region where the $X$ has been observed~\cite{alg}.    

Unfortunately, (anti)deuteron production in $pp$ collisions at $p_\perp$ values as high as $\approx 15$~GeV 
(where the $X$ is clearly seen at CMS~\cite{cms}) has not been measured yet. 

However, very recently the \alice collaboration reported results on the production of deuteron, helium-3 (\helium) 
and hypertriton (\hypertriton) light nuclei in relatively high $p_\perp$ bins in Pb-Pb 
collisions, at $\sqrt{s_\text{NN}}=2.76$~TeV~\cite{alice,alicedeuterio}. 
This is potentially a very exciting result for the reasons described above.   

We would like to draw the attention on these data and propose a way to exploit them to provide an order-of-magnitude 
estimate of light nuclei production in $pp$ collisions, to compare with the $X$ data.
\begin{figure*}[t]
 \begin{center}
   \includegraphics[width=\columnwidth]{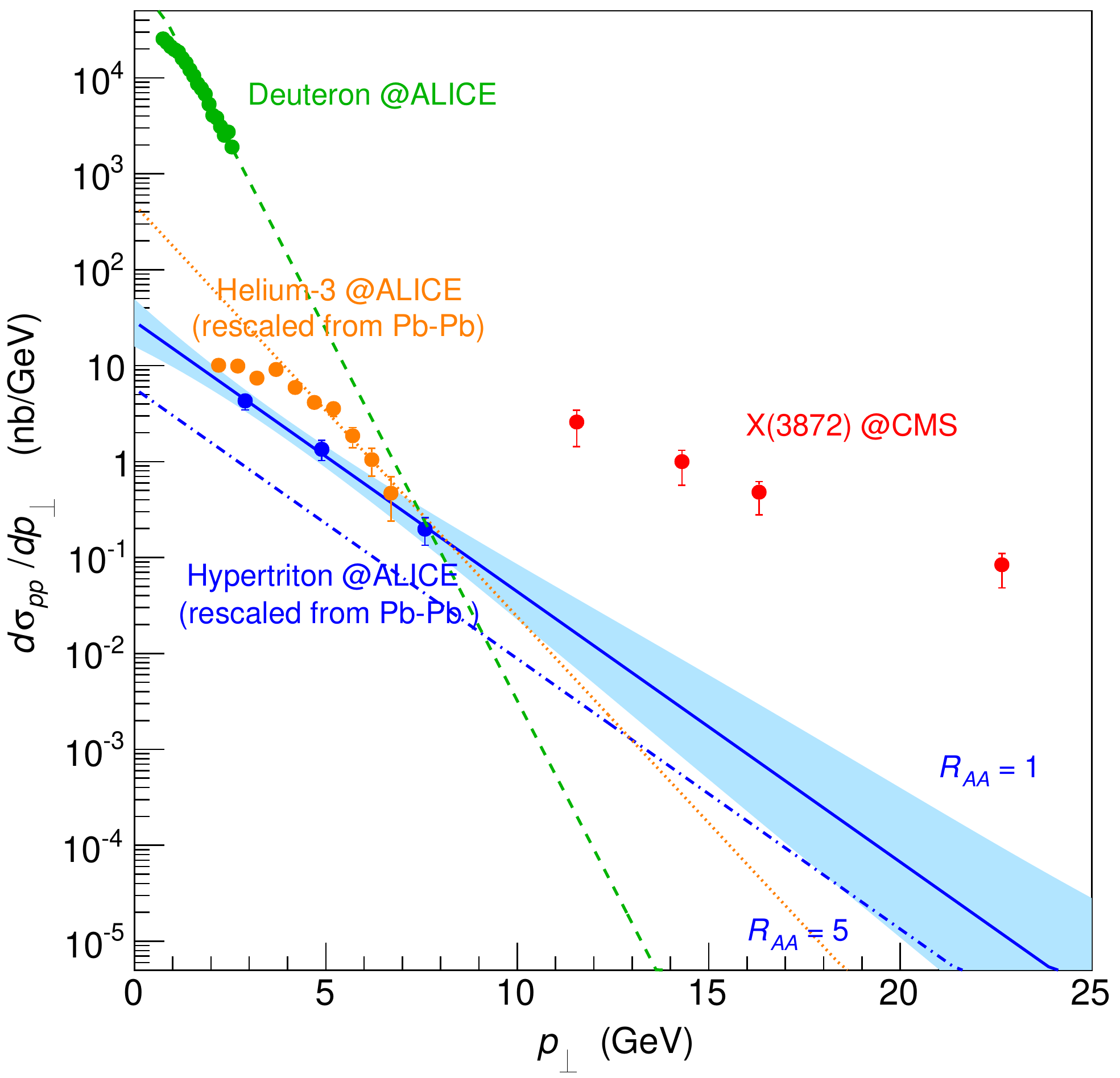} 
   \hspace{.2cm} \includegraphics[width=\columnwidth]{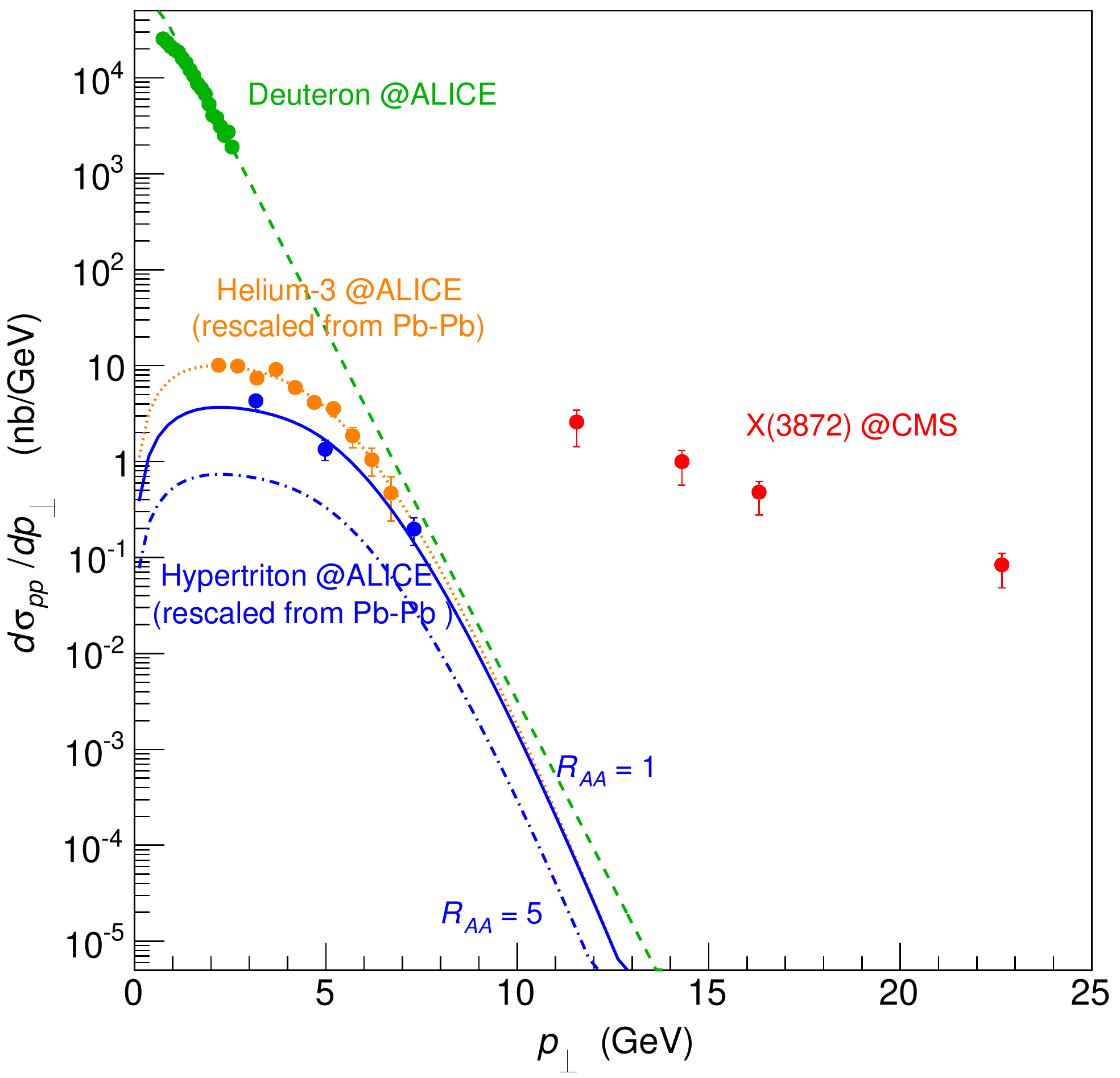}
 \end{center}
\caption{\small Comparison between the prompt production cross section in $pp$ collisions of $X(3872)$ (red), 
deuteron (green), \helium (orange), and hypertriton (blue). The $X$ data from CMS~\cite{cms} are rescaled 
by the branching ratio $\mathcal{B}(X\to J/\psi\,\pi\pi)$. Deuteron data in $pp$ collisions are taken from 
ALICE~\cite{alicedeuterio}.
The \helium and hypertriton data measured by ALICE in Pb-Pb collisions~\cite{alice,alicedeuterio} have been 
rescaled to $pp$ using a Glauber model, as explained in the text. 
The dashed green line is the exponential fit to the deuteron data points in the $p_\perp\in[1.7,3.0]$ GeV region, 
whereas the dotted orange one is the fit to the \helium data points.  The solid and dot-dashed blue lines represent 
the fits to hypertriton data with $R_{AA}=1$ (no medium effects) and an hypothetical constant value of $R_{AA}=5$. 
The hypertriton data points are horizontally shifted at the bin centres of gravity -- being defined as the point 
at which the value of the fitted function equals the mean value of the function in the bin. \label{uno}
({\bf Left Panel}) The hypertriton data are fitted with an exponential curve, and the light blue band is the 
$68\%$~C.L. for the extrapolated $R_{AA}=1$ curve. \helium data in the $p_\perp \in[4.45,6.95]$ GeV region are 
also fitted with an exponential curve. 
({\bf Right Panel}) The hypertriton and \helium data are fitted with blast-wave functions~\cite{blastwave}, 
whose parameters are locked to the \helium ones obtained in~\cite{alicedeuterio}.}
\end{figure*}

As a first approximation one can assume that there are no medium effects enhancing or suppressing the 
production of light nuclei in Pb-Pb collisions. This is equivalent to state that each nucleus-nucleus collision is 
just an independent product of $N_\text{coll}$ proton-proton collisions, with $N_{\rm coll}$ computed in a Glauber 
Monte Carlo calculation as a function of the centrality class. We use the results from~\cite{tavoleg}, which are compatible at $1\sigma$ level with the ALICE ones~\cite{alicegl}, and never more different than $3\%$. To compare with $\sqrt{s} = 7$~TeV 
data, we rescale our estimated cross sections by a factor 
$\sigma_{pp}^\text{inel}(7\text{ TeV})/\sigma_{pp}^\text{inel}(2.76\text{ TeV})=1.1$.

Consider for example the production of hypertriton observed by \alice in Pb-Pb collisions~\footnote{In the following, 
the average of hypertriton and anti-hypertriton data is understood.}. Neglecting medium effects, the $pp$ 
cross section can be estimated with 
\begin{equation}
\begin{split} \label{eq:glauber}
&\left(\frac{d\sigma\left(\hypertriton\right)}{dp_\perp}\right)_{pp}=\\
&=\quad\frac{\Delta y}{{\cal B} (\helium\,\pi)} 
\times\frac{1}{\mathcal{L}_{pp}}\left(\frac{d^2N(^3\mathrm{He}\,\pi)}{dp_\perp dy}\right)_{pp}=\\
&=\quad\frac{\Delta y}{{\cal B} (\helium\,\pi)} 
\times\frac{\sigma_{pp}^\text{inel}}{N_\text{evt}}\left(\frac{d^2N(^3\mathrm{He}\,\pi)}{dp_\perp dy}\right)_{pp}=\\
&=\quad\frac{\Delta y}{{\cal B} (\helium\,\pi)} 
\times
\frac{\sigma_{pp}^\text{inel}}{N_\text{coll}}\left(\frac{1}{N_{\text{evt}}}\frac{d^2N(\helium\,\pi)}{dp_\perp dy}
\right)_\text{Pb-Pb}.
\end{split}
\end{equation}

\alice analyzes $\helium\,\pi$ pairs,  thus we need to divide by the branching ratio for the 
$\hypertriton\to {\helium}\,\pi$ decay -- ${\cal B} (\helium\,\pi)\approx 25\%$~\cite{br} -- 
in order to deduce the number of parent hypertritons. 
We stress that the experimental data in~\cite{alice} are indeed normalized to 
$N_\text{evt}=N_\text{Pb-Pb}^\text{0-10\%}$, \emph{i.e.} the total number of inelastic Pb-Pb collisions 
analysed (about $20\times 10^6$ events in the 0-10\% centrality bin). We use $\sigma_{pp}^\text{inel}=73$~mb, 
as measured in $\sqrt{s}=7$~TeV collisions~\cite{totem}, and $\Delta y=2.4$ 
to compare with the CMS analysis~\cite{cms}. In this centrality class, we use $N_\text{coll}^\text{0-10\%} = 1518$~\cite{tavoleg}.

Similarly, we can estimate the \helium distribution in $pp$ collisions from the \alice Pb-Pb data in 
the $0$-$20\%$ centrality class~\cite{alicedeuterio}, using $N_\text{coll}^\text{0-20\%} = 1226$~\cite{tavoleg}. We remark that the selection of these events rejects any 
\helium not produced in the primary vertex, \emph{i.e.} the hypertriton decay products. 
Since the \helium data points with $p_\perp < 4.4$~GeV show a deviation from the exponential behavior, 
likely due to the expansion of the medium, we perform an exponential fit to the points in the region 
$p_\perp \in[4.45,6.95]$~GeV only. Alternatively, we fit hypertriton and \helium data with the blast-wave 
model~\footnote{The blast-wave function is 
\begin{equation*}
\frac{dN }{dp_\perp}\propto p_\perp \int_0^R r dr \,m_\perp I_0\!\left(\frac{p_\perp \sinh \rho }{ T_\text{kin}} 
\right) K_1\!\left(\frac{m_\perp \cosh \rho }{ T_\text{kin}} \right),
\end{equation*} where $m_\perp$ is the transverse mass, $R$ is the radius of the fireball, $I_0$ and $K_1$ 
are the Bessel functions, 
\mbox{$\rho = \tanh^{-1} \left(\frac{(n+2)\left\langle\beta\right\rangle}{2} (r/R)^n\right)$}, 
and $\left\langle\beta\right\rangle$ the averaged speed of the particles in the medium. }, 
which describes particle production properties by assuming thermal emission from an expanding 
source~\cite{blastwave}. This model is expected to reproduce correctly the low and medium $p_\perp$ 
regions in Pb-Pb collisions. Since we are rescaling Pb-Pb data to $pp$ by a constant factor, the same shape 
holds in our estimated $pp$ data, and gives a guess on the asymptotic exponential behavior. 
The results are shown in Fig.~\ref{uno}.

Our rescaling to $pp$ collisions does not  take into account neither medium  effects, nor the fact that the 
coalescence/recombination mechanism can be enhanced in Pb-Pb collisions~\cite{Sato:1981ez}. In fact, such 
phenomena are known to favor the production of many-body hadrons with respect to what expected in vacuum. 
Medium effects are discussed later. 
 
 \begin{figure}[t]
 \begin{center}
   \includegraphics[width=.45\textwidth]{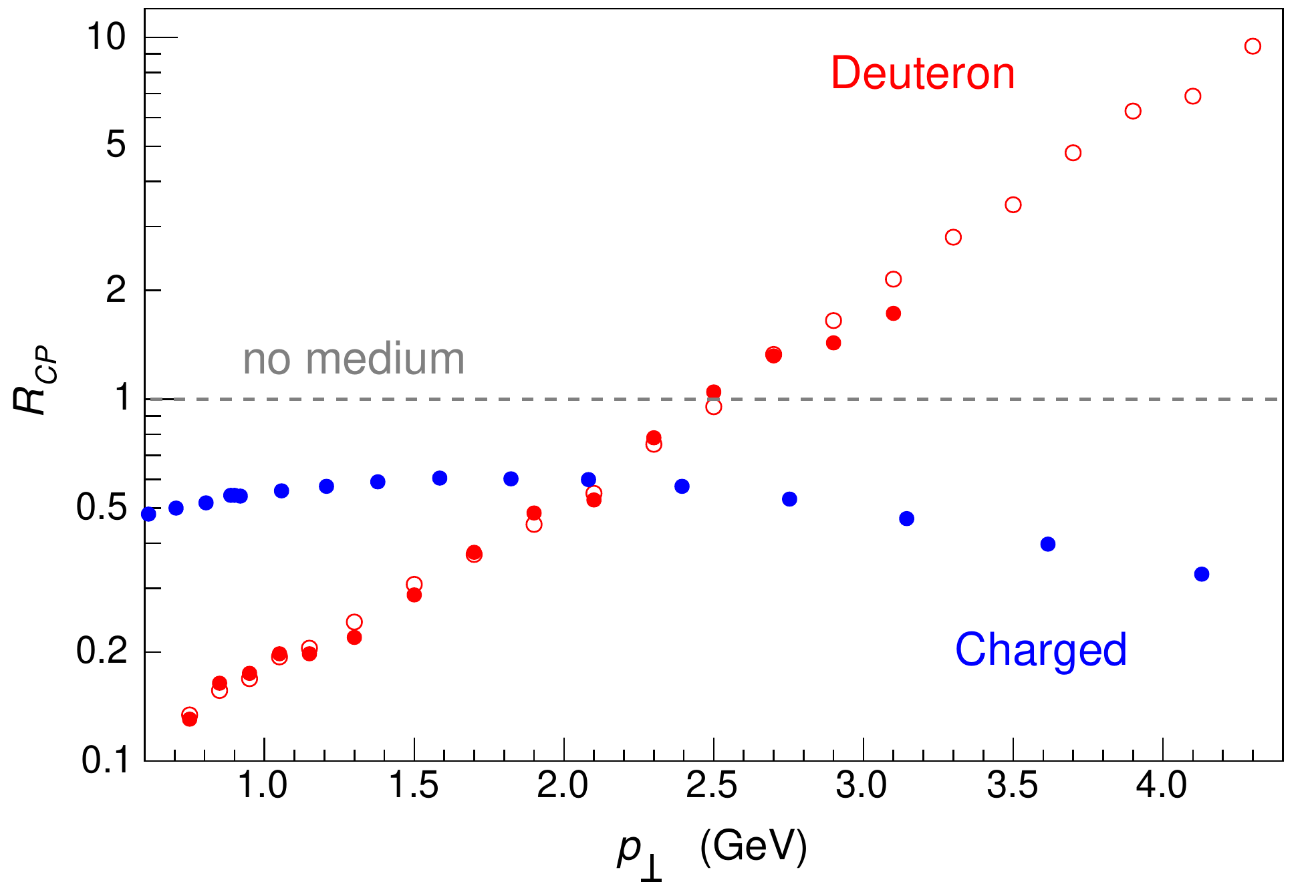}
 \end{center}
\caption{\small 
Comparison between the nuclear modification factor $R_{CP}$ for deuteron (red) and for generic charged tracks 
(blue)~\cite{raaatlas} in central (resp. $0$-$10\%$ and $0$-$5\%$) versus peripheral ($60$-$80\%$) Pb-Pb collisions. 
We evaluate $R_{CP}$ either taking the bin-by-bin ratio of ALICE data~\cite{alicedeuterio} (full disks), 
or using the values given by the blast-wave fit for the peripheral Pb-Pb data, and extrapolated up to 
$p_\perp \sim 4.8$~GeV (open disks).
The dashed line corresponds to no medium effects, $R_{CP}=1$. It is worth noticing that  $R_{CP}(d)$ 
gets enhanced at \mbox{$p_\perp \gtrsim 2.5$~GeV}. \label{due}}
\end{figure}

For the deuteron we use \alice $pp$ data~\cite{alicedeuterio} to estimate 
\begin{equation}
\left(\frac{d\sigma\left(d\right)}{dp_\perp}\right)_{pp}=\Delta y 
\times \sigma_{pp}^\text{inel}\left(\frac{1}{N_{pp}^\text{inel}}\frac{d^2N(d)}{dp_\perp dy}\right)_{pp}
\end{equation}
$N^\text{inel}_{pp}$ being the number of $pp$ inelastic collisions collected. 
We perform the fit to the points in the region $p_\perp \in[1.7,3.0]$~GeV, which shows a good exponential behaviour.

The CMS analysis of $X$ production provides the differential cross section times the branching 
fraction $\mathcal{B}\left(X(3872) \to J/\psi\,\pi^+\pi^-\right)$. The latter has not been measured yet, 
and the lower limit reported in the PDG is $\mathcal{B} > 2.6\%$~\cite{pdg}. An estimate for the upper limit 
has been reported, $\mathcal{B} < 6.6\%$ at 90\% C.L.~\cite{yuanx};
we use instead the more conservative value $\mathcal{B} = 8.1^{+1.9}_{-3.1}\%$~\cite{review}.  
The comparison in Fig.~\ref{uno} shows that, according to the most conservative exponential 
fit in the left panel, the extrapolated hypertriton production cross section in $pp$ collisions would 
fall short by about $2\div 3$ orders of magnitude with respect to the $X$ production, and much more 
according to the blast-wave fit in the right panel. The drop of the deuteron cross section, 
which is directly measured in $pp$ collisions, appears definitely faster. 

As we  mentioned already, the main problem for the production of loosely bound molecular states in 
proton-proton collisions is the difficulty in  producing the constituents close enough in phase space. 
However, it is well known that the interaction of elementary partons with the collective hot dense 
medium causes relevant energy loss of the partons themselves. This effect is usually quantified by 
the nuclear modification factor~\cite{raaphenix,raastar,raaatlas,raacms,aliceraa} 
\begin{equation}
R_{AA}=\frac{\left(\frac{1}{N_\text{evt}}\frac{d^2N}{dp_\perp dy}\right)_\text{Pb-Pb}}{N_\text{coll} 
\left(\frac{1}{N_\text{evt}} \frac{d^2N}{dp_\perp dy}\right)_{pp}},
\end{equation}
which compares the particle yield in Pb-Pb collisions with that in $pp$. It then follows that the method used 
to obtain Eq.~(\ref{eq:glauber}) corresponds  to assume $R_{AA}=1$.

While for ordinary hadrons medium effects generally lead to a suppression of the particle yield -- \emph{i.e.} 
$R_{AA}<1$ -- conversely they can favor the production of hadronic molecules. The role of the medium would be, 
in fact, that of decreasing the relative momenta of the components with respect to the zero temperature case 
due to the well-known jet quenching effect~\cite{Gyulassy,Baier:1994bd}. This would favor their coalescence 
into the final bound state by reducing their relative momenta directly at parton level.

The coalescence model is based on the sudden approximation~\footnote{\emph{i.e.} the assumption that the 
binding of the constituents happens on small time scales and therefore their wave function remains unchanged 
during the transition to the bound state.} and is implemented by calculating the overlap of the density 
matrix of the constituents with the Wigner function of the final composite particle. In particular, it has 
the important property of taking into account the inner structure of the considered hadron. If one only 
requires vicinity in momentum space, the $p_\perp$ distribution of a composite state with $N$ constituents 
coming out of a hot QCD medium is roughly given by
\begin{align}
\frac{dN_\text{b}}{dp_\perp}(\bm p_\perp)\sim\prod_{i=1}^N\frac{dN_i}{dp_\perp}(\bm p_\perp/N),
\end{align}
where $N_\text{b}$ is the number of final bound states and $N_i$ is the number of produced constituents. 
This would also explain why in Fig.~\ref{uno} the cross section for the \helium and hypertriton are several 
orders of magnitude smaller than the deuteron one: one additional $p$ or $\Lambda$, close enough in phase space, 
must be produced.

 It has already been shown that coalescence effects in Pb-Pb collisions can have relevant consequences on the 
production of multi-quark states. In particular, molecular states with small binding energy are expected to 
be \emph{enhanced}, \emph{i.e.} $R_{AA}>1$~\cite{Cho:2010db}.

Unfortunately there is no measurement of $R_{AA}$ for the deuteron as a function of $p_\perp$.
However, there is another nuclear modification factor which is often used,
\begin{equation}
R_{CP}=\frac{\left(\frac{1}{N_\text{evt}}\frac{d^2N}{dp_\perp dy}\right)_\text{Pb-Pb}^\text{0-10\%} \Big{/} 
N_\text{coll}^\text{0-10\%} }{ \left(\frac{1}{N_\text{evt}} \frac{d^2N}{dp_\perp dy}\right)_\text{Pb-Pb}^\text{60-80\%} 
\Big{/} N_\text{coll}^\text{60-80\%}}.
\end{equation}
This quantity is a comparison between the most central and the most peripheral Pb-Pb collisions and therefore 
provides another valid indicator of the strength of medium effects (which should be absent in the less dense, 
most peripheral events). The fact that  $R_{AA}$ and $R_{CP}$ measurements for hadron species are strongly 
correlated to each other is shown experimentally by a thorough data analysis reported
by ATLAS~\cite{raaatlas}, up to very high $p_\perp \sim 100$~GeV.

Using the \alice data presented in~\cite{alice} we can compute $R_{CP}$ for deuteron as a function of $p_\perp$ 
and compare it with that for generic charged tracks, as reported in~\cite{raaatlas} -- see Fig.~\ref{due}. We use $N_\text{coll}^\text{60-80\%} = 27.5$~\cite{tavoleg}.
As one immediately notices, the difference from ordinary hadrons is striking. The presence of the QCD medium 
is extremely effective at enhancing the production of deuteron for the reasons explained before. In fact, 
$R_{CP}$ for this hadronic molecule becomes larger than unity for $p_\perp \gtrsim 2.5$~GeV, in particular 
we have $R_{CP} = 1.7$ at the last point with $p_\perp = 3.1$~GeV. Using the blast-wave fitting function 
for the peripheral data taken from~\cite{alice}, we also extrapolate up to the end point of the central data, 
confirming the growth of $R_{CP}$ with $p_\perp$. 

We expect a similar behavior in $R_{AA}$, in particular a value larger than 1 for $p_\perp$ large enough. 

To get an independent rough estimate for $R_{AA}$, we assume the deuteron production cross section in $pp$ 
collisions to scale with $\sqrt{s}$ like the inelastic cross section, and compare the ALICE data in central 
Pb-Pb collisions at $\sqrt{s_\text{NN}}=2.76$~TeV, with the ones in $pp$ 
collisions at~$\sqrt{s}=7$~TeV~\cite{alicedeuterio}. Indeed, we find that $R_{AA}$ exceeds 1 at 
$p_\perp = 2.1$~GeV, and reaches $5$ at $p_\perp = 4.3$~GeV. This gives strength to our expectation 
for $R_{AA} > 1$. To display the size of this effect, we plot also the hypertriton curves for 
$R_{AA} = 5$ in Fig.~\ref{uno}.

One naturally expects for a similar enhancement to be even more relevant for $3$-body nuclei like \helium 
and hypertriton. Its role would be to further \emph{decrease} the extrapolated cross section in prompt 
$pp$ collisions. As we already said, indeed, a value of $R_{AA} > 1$ applied to Pb-Pb data implies a $pp$ cross section even smaller than predicted by the Glauber model. 
Even though qualitative conclusions can already be drawn, a quantitative analysis 
substantiated by data at higher $p_\perp$ is necessary for a definitive comparison with the $X$ case. 

Even assuming that only a hot pion gas is excited in  Pb-Pb collisions, there would likely be a large number 
of final state interactions with pions catalizing the formation of a loosely bound hypertriton along the lines 
discussed in~\cite{angetc,alg,review}. In any case, such an environment is present in the Hadron Resonance Gas 
corona formed when the outer shell of the QCD medium cools down~\cite{corona}.

In summary, the extrapolation of deuteron and \mbox{\helium} data in $pp$ collisions shown in Fig.~\ref{uno} 
suggests that loosely bound molecules are hardly produced at high $p_\perp$. The extrapolated curve of 
hypertriton data from Pb-Pb collisions might lead to milder conclusions although we expect it should  
be significantly suppressed when medium effects are properly subtracted. Such effects are indeed already sizeable 
for the deuteron as shown in Fig.~\ref{due}, and probably even more relevant for $3$-body nuclei.

We are aware that for an unbiased and definitive comparison with $X$ production at $p_\perp$ as high as 15~GeV, 
deuteron (or hypertriton) should be searched in $pp$ collisions rather than in Pb-Pb to avoid the complicacies 
of subtracting medium effects. These analyses can be performed by \alice and LHCb during Run~II. One of the 
purposes of this letter is to further motivate the required experimental work.

\textbf{Acknowledgments:} We wish to thank M.~Gyulassy for valueable discussions on medium effects and 
C.~Hanhart for pointing the ALICE results to us. We also thank M.A.~Mazzoni for interesting comments and 
suggestions, and S.~Bufalino, A.~Caliv\`a, G.~Cavoto, B.~Doenigus, A.P.~Kalweit, R.~Lea for clarifications on
data published by the ALICE collaboration. 
%


\end{document}